\documentclass[%
 reprint,
 amsmath,amssymb,
 aps,
]{revtex4-1}

\usepackage{graphicx}% Include figure files
\usepackage{dcolumn}% Align table columns on decimal point
\usepackage{ulem}
\usepackage{bm}% bold math
\usepackage[dvipsnames]{xcolor}

\newcommand{\done}[1]{\textcolor{black}{#1}}
\newcommand{\revtwo}[1]{\textcolor{black}{#1}}

\usepackage{hyperref}% add hypertext capabilities
\usepackage{tikz}

\begin{document}

\preprint{APS/123-QED}

\title{Flow Coupling Alters 
Topological Phase Transition in Nematic Liquid Crystals}% Force line breaks with \\
%\thanks{A footnote to the article title}%

\author{Jayeeta Chattopadhyay}
\email{These authors contributed equally to this work}
\author{Simon Guldager Andersen}
\email{These authors contributed equally to this work}
\author{Kristian Thijssen}%
\email{kristian.thijssen@nbi.ku.dk}
\author{Amin Doostmohammadi*}%
\email{doostmohammadi@nbi.ku.dk\\ \noindent * primary corresponding author}
\affiliation{Niels Bohr Institute, University of Copenhagen, Blegdamsvej 17, Copenhagen, Denmark.}

\date{\today}% It is always \today, today,
             %  but any date may be explicitly specified
\begin{abstract}
We investigate how coupling to fluid flow influences defect-mediated transitions in two-dimensional passive \done{and active} nematic fluids using fluctuating nematohydrodynamic simulations. The system is driven by tuning the fluctuation strength, with increasing (decreasing) fluctuations defining the forward (backward) protocol \done{through the defect creation temperature threshold where defects spontaneously nucleate.} In the absence of flow coupling, the transition follows the Berezinskii--Kosterlitz--Thouless (BKT) scenario, governed by reversible binding and unbinding of $\pm 1/2$ defect pairs. When hydrodynamics is included, the outcome is controlled by the flow--alignment parameter. For non-aligning nematics ($\lambda=0$), the transition remains consistent with BKT. By contrast, for strain-rate--\done{coupled} nematics ($\lambda\neq 0$), bend--splay walls emerge, lowering the defect nucleation threshold and preventing sustained recombination: once created, defects remain unbound across the full range of fluctuation strengths in both forward and backward protocols. \done{In active nematics,
self-generated flows produce the same effect, with defects remaining unbound irrespective of alignment.
These findings demonstrate that coupling to flow, whether through strain-rate alignment or
activity, fundamentally alters defect-mediated phase transitions and suggest that the canonical BKT transition emerges only in the absence of flow coupling.}
%These results identify flow alignment as a fundamental control parameter for topological phase behavior and suggest that the canonical BKT transition emerges only in the absence of flow alignment. 

%\vspace{1em}
%\sg{NOTE: Now that we explicitly show that our results are independent of system size, we might want to make a big deal out of it and mention it in some places}

%\jc{Comment: Unbinding and defect nucleation temperatures are different. Do we really need to add the red-marked sentence to the abstract?}.

%\kt{Regarding Jayeetas comment. I have modified it a tiny bit. you are right it doesn't add anything, BUT we need somethingt to add something to address the referees point, and because his point is useless, the addition to the abstract will be useless. But we need to have something to keep him happy.}
\end{abstract}

%\keywords{Suggested keywords}%Use showkeys class option if keyword
                              %display desired
\maketitle

\section{Introduction}
Topological defects are fundamental excitations in two-dimensional nematic fluids. As singularities of the orientational order parameter, they impact both the stability of ordered phases and the mechanisms of disordering in passive \cite{de1993physics, chuang1991, bowick1994} as well as active nematics \cite{shankar-prl-2018, andersen2025, shankar-prx-2019, thampi_active_2016,ruider2025topological,carenza2025quasi}.
In purely relaxational dynamics, where nematics are uncoupled from hydrodynamic flow, the topological phase transition follows the Berezinskii–Kosterlitz–Thouless (BKT) scenario \cite{berezinskii1970, kosterlitz1973, kosterlitz1974, stein1978kosterlitz}. In this framework, defects of charge $\pm 1/2$ form bound pairs at low temperature, leading to quasi–long-range nematic order. Increasing temperature drives the unbinding of oppositely charged pairs, which proliferate and ultimately destroy orientational order \cite{kosterlitz2018ordering, stein1978kosterlitz, KT-LC-2014}. This transition has been established both analytically, through renormalization group theory \cite{domb2000phase}, and computationally~\cite{bates2000phase}.

Coupling the nematic orientation field to hydrodynamic flow significantly enriches this picture. \done{Hydrodynamic backflow induces a marked asymmetry in defect dynamics: $+1/2$ defects are advected and accelerated by the flow field they generate, whereas $-1/2$ defects remain comparatively immobile} \cite{toth-prl, Blac-prl, biscari2012perturbative}. This kinematic anisotropy alters defect–defect interactions and, in turn, the collective behavior of defect ensembles. A key question, therefore, is whether such flow coupling merely renormalizes the BKT transition temperature or whether it can fundamentally modify the nature of the transition itself.

The coupling of nematics to hydrodynamic flows crucially depends on the response of nematogens to velocity gradients, governed by the flow–alignment parameter, which quantifies whether the director tends to align with or tumble in a shear flow. This parameter is nonzero in most experimental systems and thus represents a generic feature of \done{nematic materials \cite{PhysRevE.86.041701,krekhov2000nematic,PhysRevLett.28.1554,PhysRevLett.32.924}}. Its role in shaping defect dynamics and phase transitions, however, remains insufficiently understood. In particular, it is not known whether flow alignment preserves the BKT binding–unbinding mechanism or whether it can destabilize bound pairs altogether.

In this work, we address this question by performing large-scale two-dimensional fluctuating nematohydrodynamic simulations. We first recover the BKT transition in the absence of hydrodynamic coupling, confirming the binding–unbinding mechanism of $\pm 1/2$ defect pairs. Upon including hydrodynamics, we demonstrate that the nature of the transition depends crucially on the flow–alignment response: while non-aligning nematics remain consistent with the BKT picture, \done{strain-rate coupled} nematics display persistent defect unbinding across all temperatures.
\done{In active nematics, self-generated flows also eliminate defect binding, independent of flow alignment.
These results identify flow coupling as a fundamental control parameter for defect-mediated transitions, showing how activity and hydrodynamics reshape topological phase behavior in nematic fluids.}
%These results identify the flow–alignment parameter as a fundamental control variable of defect-mediated transitions in nematic fluids, showing how hydrodynamics can reshape topological phase behavior.
%
\begin{figure*}[t!]
    \centering
\includegraphics[height=0.9\linewidth, width=\linewidth]{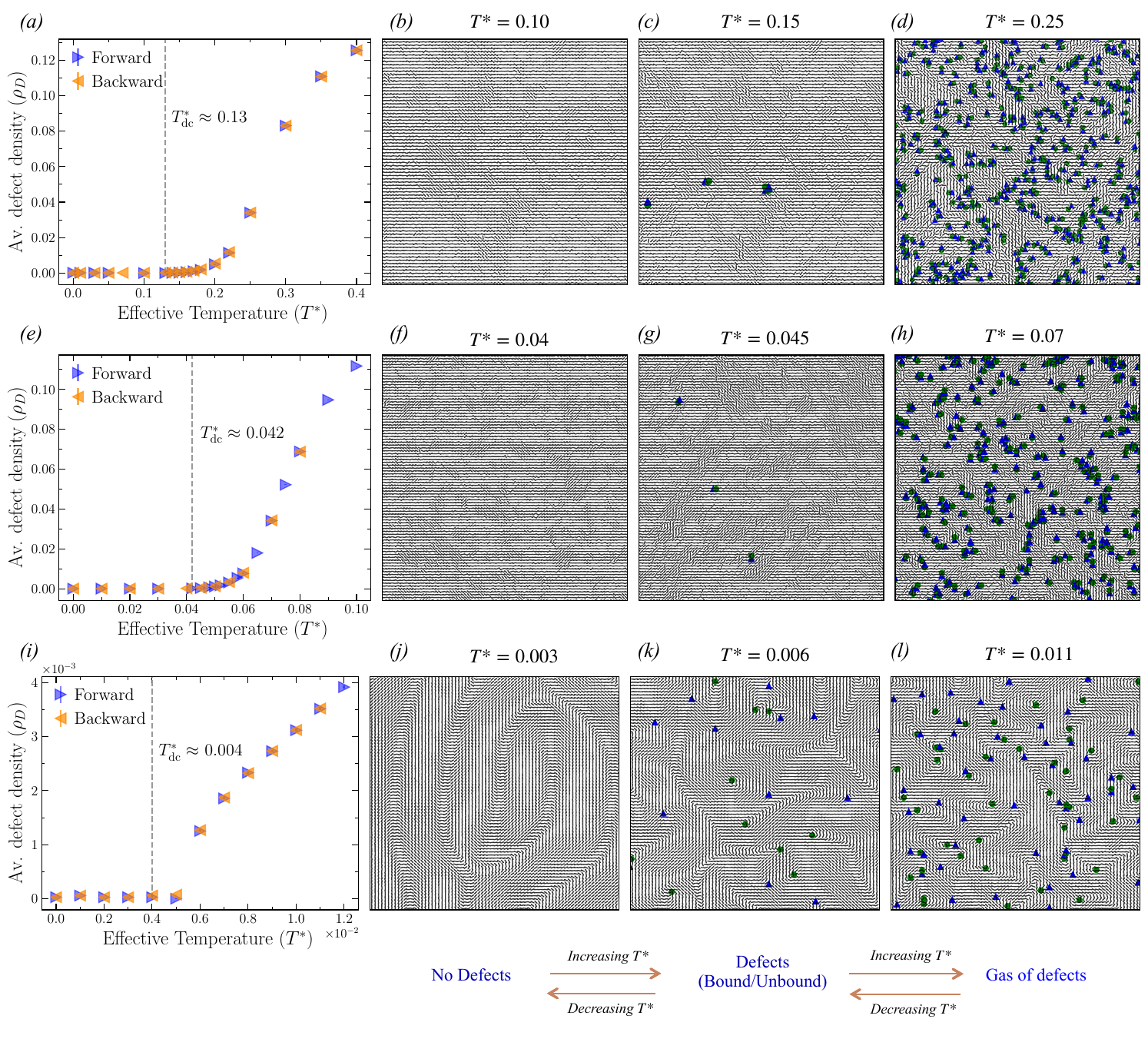}

%  \caption{The transition from defect-free to defect-laden states under temperature fluctuations, with snapshots of equilibrium configurations for passive nematic systems: (a–d) without flow coupling, and (e–l) with hydrodynamic coupling. The latter includes (e–h) non-aligning nematics with $\lambda = 0$ and (i–l) strain-rate–aligning nematics with $\lambda = 1$. 
% Vertical dashed lines indicate the defect creation temperatures \done{$T^*_{dc}$} for each case. Black lines represent the director field, while $+1/2$ ($-1/2$) defects are marked by green circles (blue triangles). 
% For flow-decoupled as well as non-aligning nematics, increasing the temperature (forward protocol) drives the system through successive transitions: from (b,f) a defect-free state, to (c,g) a state of bound defect pairs, and finally to (d,h) a defect gas with numerous unbound defects. Conversely, decreasing the temperature (backward protocol) from (d,h) results in the (c,g) rebinding and annihilation of defects, eventually restoring (b,f) a defect-free state .
% In contrast, for strain-rate–aligning nematics ($\lambda = 1$), the system evolves from (j) a defect-free state to (k) a state where defect pairs unbind immediately upon creation, resulting in unbound defects, and finally to (l) a defect-rich state with numerous free defects as the temperature increases. Lowering the temperature retraces the same path, so defects remain unbound whenever present.}
\caption{\done{Transition from defect-free to defect-laden states under \done{increasing effective} temperature in passive nematics. Panels (a–d) show systems without flow coupling; panels (e–l) include hydrodynamic coupling, with (e–h) non-aligning nematics ($\lambda=0$) and (i–l) strain-rate–aligning nematics ($\lambda=1$). 
\done{In panels (a, e, i), the average defect density $\rho_D$ is plotted on the y-axis as a function of the effective temperature $T^*$. Vertical dashed lines mark the defect-creation temperature $T^*_{dc}$ defined as the temperature at which $\rho_D$ becomes nonzero.} 
Black lines indicate the director field, with $+1/2$ and $-1/2$ defects shown as green circles and blue triangles, respectively.
For flow-decoupled and non-aligning nematics, increasing temperature drives the system from a defect-free state (b,f) to one of bound defect pairs (c,g) and finally to a defect gas of unbound defects (d,h); cooling reverses this path with pair rebinding and annihilation. In contrast, strain-rate–aligning nematics evolve from a defect-free state (j) directly to one of unbound defect pairs (k,l); 
upon cooling, defects remain unbound.}}
\label{Fig1:unbinding_schematic}
\end{figure*}
\section{Model}
We use a two-dimensional fluctuating nematohydrodynamic model 
\cite{doostmohammadi2018active,thampi2016active} 
that is convenient for (de-) coupling the collective flow field and orientational field \cite{doostmohammadi2016stabilization}. 
The nematic order parameter tensor $\mathbf{Q}$ evolves according to the Beris--Edwards equations \cite{beris_thermodynamics_1994}:
\begin{equation}
\partial_t \mathbf{Q} + \vec{u} \cdot \vec \nabla \mathbf{Q} - \mathbf{S} = \Gamma\mathbf{H} + \mathbf{\xi}^{Q} .
\label{eq-Q}
\end{equation}

The velocity field $\vec{u}$ evolves according to the generalized 
Navier--Stokes equations for an incompressible fluid:
\begin{equation}
\rho \left( \partial_t \vec{u} + \vec{u} \cdot \vec \nabla \vec{u} \right) 
= \vec\nabla \cdot \mathbf{\Pi} + \vec \nabla \cdot \mathbf{\xi}^u, 
\quad \vec \nabla \cdot \vec{u} = 0.
\label{eq-u}
\end{equation}

In Eq.~\ref{eq-Q}, 
\begin{equation}
    \mathbf{H} = -\frac{\partial F}{\partial \mathbf{Q}} + \frac{\mathbf{I}}{2} Tr\left (\frac{\partial F}{\partial \mathbf{Q}} \right ),
\end{equation}
denotes the molecular field, describing relaxation 
towards the minima of the free energy $F$, which contains both 
the bulk Landau–de Gennes contribution and the elastic Oseen--Frank term
\begin{equation}
    F = A \left(1 - \frac{1}{2} \text{Tr}(\mathbf{Q}^2)\right)^2 + \frac{K}{2} (\vec{\nabla} \mathbf{Q})^2.
\end{equation}
The bulk parameters are chosen such that the system is deep within the nematic state, and local nematic order remains high. 
%Hence, in order for defects to nucleate, they require a certain amount of energy to overcome this energy barrier 
The parameters $\rho$ and $\Gamma$ are the fluid density and rotational diffusivity, respectively. 
$\mathbf{S}$ is the co-rotational advection term accounting for the interplay between flow and particle orientation via the flow-alignment parameter $\lambda$:  
\begin{equation}
\begin{aligned}
\mathbf{S} &= (\lambda \mathbf{E} + \mathbf{\Omega}) \cdot \left(\mathbf{Q} + \tfrac{\mathbf{I}}{2}\right) 
+ \left(\mathbf{Q} + \tfrac{\mathbf{I}}{2}\right) \cdot (\lambda \mathbf{E} - \mathbf{\Omega}) \\
&\quad - 2\lambda \left(\mathbf{Q} + \tfrac{\mathbf{I}}{2}\right)(\mathbf{Q} : \vec{\nabla} \mathbf{u}),
\end{aligned}
\end{equation}
where $\mathbf{E}$ is the strain rate, 
$E_{ij} = \tfrac{1}{2} (\partial_i u_j + \partial_j u_i)$, 
and $\mathbf{\Omega}$ is the vorticity tensor, 
$\Omega_{ij} = \tfrac{1}{2} (\partial_i u_j - \partial_j u_i)$. 
$\lambda$ couples the orientation to strain rates of the flow. $\lambda = 0$ corresponds to non-aligning nematics, where the director is affected only by vortical flows and does not respond to the strain rate, whereas $\lambda = 1$ represents strain-rate–aligning nematics, where the director tends to align with shear flow \cite{edwards2009spontaneous}. 

In Eq.~\ref{eq-u}, the stress tensor $\mathbf{\Pi}$ includes pressure, viscous, and elastic contributions: 
$\Pi_{ij}^{\text{pressure}} = -p\delta_{ij}$,  
$\Pi_{ij}^{\text{viscous}} = 2\eta E_{ij}$ with dynamic viscosity $\eta$,  
and an elastic part, that accounts for backflow \cite{kos2020field, toth-prl, Blac-prl}: $\Pi^{\text{elastic}}_{ij} = 
2 \lambda \left(Q_{ij} + \tfrac{\delta_{ij}}{2}\right) 
   \left(Q_{lk} H_{kl}\right)
- \lambda H_{ik} \left(Q_{kj} + \tfrac{\delta_{kj}}{2}\right)
- \lambda \left(Q_{ik} + \tfrac{\delta_{ik}}{2}\right) H_{kj}
- \partial_i Q_{kl} \, \frac{\delta F}{\delta (\partial_j Q_{lk})}
+ Q_{ik} H_{kj} - H_{ik} Q_{kj}$.  
%The parameter $\gamma$ in Eq.~\ref{eq-u} represents substrate friction.  
%For small $\gamma$, viscous forces dominate and flows propagate through the system. 
%In the opposite limit of large $\gamma$, momentum is strongly damped by friction and the dynamics is overdamped. 
%In this case, Eq.~\ref{eq-u} reduces to
%\begin{equation}
%\gamma \vec{u} \approx - \zeta \vec \nabla \cdot \mathbf{Q} + \vec \nabla \cdot \mathbf{\xi}^u .
%\end{equation}
%and Eq.~\ref{{eq-Q}} becomes a function of only $Q$. 
%Throughout this work, we refer to these limits as viscous ($\gamma=0$) and friction-dominated ($\gamma=1$).

%%%%
We model thermal fluctuations in the nematic order and flow fields, denoted $\mathbf{\xi}^Q$ and $\mathbf{\xi}^u$, as Gaussian random variables with zero mean and variance set by \cite{Lasse-PRE}:
\begin{equation}
\langle \xi_{ij}^{Q}(\vec{x}, t) \xi_{kl}^{Q}(\vec{x}', t') \rangle 
= 2k_B T^Q \Gamma J_{ijkl} \delta(\vec{x} - \vec{x}') \delta(t - t'),
\end{equation}
\begin{equation}
\langle \xi_{ij}^{u}(\vec{x}, t) \xi_{kl}^{u}(\vec{x}', t') \rangle 
= 2k_B T^u \eta J_{ijkl} \delta(\vec{x} - \vec{x}') \delta(t - t'),
\end{equation}
% where $J_{ijkl} = \delta_{ik}\delta_{jl} + \delta_{il}\delta_{jk} - \delta_{ij}\delta_{lk}$ ensures that $\mathbf{Q}$ remains traceless and symmetric~\cite{bertin2013mesoscopic}.  
 %%%%%
% The fluctuation strengths are expressed as $\hat{Q}_{k_B T} = k_B T^Q / K$ and $\hat{u}_{k_B T} = k_B T^u / K$, normalized by the elastic constant $K$ \done{(which has units of energy in 2D) to enable} direct comparison with experiments \cite{duclos2017topological}. 
% Here, \(T^Q\) and \(T^u\) denote the effective temperatures proportional to the orientational and velocity fluctuations, respectively. 
% When these fluctuations are balanced such that $T^{Q}=T^{u}$, and the correlations follow the form given above, i.e. 
% behave according to the fluctuation-dissipation theorem for a fluid at temperature $T^{Q}=T^{u}$~\cite{hohenberg1977, adhikari2005}, detailed balance is preserved
% ~\cite{Lasse-PRE, thampi_lattice-boltzmann-langevin_2011}. In other words, the stochastic terms do not introduce net driving forces beyond those arising from dissipation and elastic restoring forces, thereby maintaining time‐reversal symmetry.

% \done{Numerical values are given in table \ref{table:simulation_parameters}. The values are in similar regime to previous works on nematics \cite{thampi2014vorticity,shendruk2017dancing,Lasse-PRE}. The unbinding temperature is set by the free energy constants, with different parameters just normalizing the Ericksen number \cite{fedorowicz2023effects} without affecting the qualitative physics presented in this manuscript. }
%%%%%
\noindent
where $J_{ijkl} = \delta_{ik}\delta_{jl} + \delta_{il}\delta_{jk} - \delta_{ij}\delta_{lk}$ ensures that $\mathbf{Q}$ remains traceless and symmetric~\cite{bertin2013mesoscopic}, \done{and $k_B$ is the Boltzmann constant. 
The strengths of fluctuation are set by \(T^Q\) and \(T^u\), which have units of temperature, and when fluctuations in $\textbf{Q}$ and $\vec{u}$ are balanced in the sense} that $T^{Q}=T^{u}$, detailed balance is preserved
~\cite{hohenberg1977,adhikari2005, Lasse-PRE,thampi_lattice-boltzmann-langevin_2011}. In other words, \done{the stochastic terms involving $\mathbf{\xi}^Q$ and $\mathbf{\xi}^u$ in Eqs. \eqref{eq-Q} \& \eqref{eq-u}} do not introduce net driving forces beyond those arising from dissipation and elastic restoring forces, thereby maintaining time‐reversal symmetry.
% The time-reversal symmetry of the Beris-Edwards and Navier-Stokes equations, is maintained 
% adding stochastic terms ($\mathbf{\xi^Q}$ and $\mathbf{\nabla \cdot \mathbf{\xi}^u}$)
% to the Beris-Edwards and Navier-Stokes equations, as we have done in Eqs. \eqref{eq-Q} \& \eqref{eq-u}, 
\done{
In light of this, we set $T^{Q}=T^{u} \equiv T$ and define a single, dimensionless control parameter for the fluctuation strengths as $T^* = k_B T / A $, which we shall refer to as \textit{the effective temperature}.}

\done{ Eqs. \eqref{eq-Q} and \eqref{eq-u} have been solved numerically using the hybrid Lattice Boltzmann method for square systems with side length $L\in \{128, 256,512\}$, periodic boundary conditions, and a range of effective temperatures (see App. \ref{app:hb} for details).}
\done{
Model parameters are chosen in similar regime to previous works on nematics \cite{thampi2014vorticity,shendruk2017dancing,Lasse-PRE}, and unless otherwise stated, all parameters and numeric values are expressed in units of the lattice spacing $\Delta x = 1$, the simulation time step $\Delta t = 1$, and the Landau coefficient $A = 1$ (Tab. \ref{table:simulation_parameters}).}
\done{
\done{The temperature where defects nucleate} is set by the free energy constants, \done{with different parameters shifts the Ericksen number} \cite{fedorowicz2023effects} without affecting the qualitative physics presented in this manuscript.}
% Here, \(T^Q\) and \(T^u\) denote the effective temperatures proportional to the orientational and velocity fluctuations, respectively. 
% When these fluctuations are balanced such that $T^{Q}=T^{u}$, and the correlations follow the form given above, i.e. 
% behave according to the fluctuation-dissipation theorem for a fluid at temperature $T^{Q}=T^{u}$~\cite{hohenberg1977, adhikari2005}, detailed balance is preserved
% ~\cite{Lasse-PRE, thampi_lattice-boltzmann-langevin_2011}. In other words, the stochastic terms do not introduce net driving forces beyond those arising from dissipation and elastic restoring forces, thereby maintaining time‐reversal symmetry.
% Model parameters are in similar regime to previous works on nematics \cite{thampi2014vorticity,shendruk2017dancing,Lasse-PRE}. The unbinding temperature is set by the free energy constant, with different parameters just normalizing the Ericksen number \cite{fedorowicz2023effects} without affecting the qualitative physics presented in this manuscript.
% Unless otherwise stated, all parameters are expressed in units of the lattice spacing $\Delta x = 1$, the simulation time step $\Delta t = 1$, and the Landau coefficient $A = 1$ (Tab. \ref{table:simulation_parameters}).
% MENTION: results are for L=256. with that one fig in appendix to show that results are ind of syystem size.
%-----------------------------------------------------------------------------
\begin{table}[h!]
\centering
\caption{\done{Values/ranges and dimension of simulation parameters. $L, M, T$ refer to length, mass and time, respectively. All parameters are expressed in units of ($\Delta x$, $\Delta t$, $A$).}}
\resizebox{\columnwidth}{!}{
\begin{tabular}{|c|c|c|c|}
\hline
\textbf{Parameter} & \textbf{Symbol} & \textbf{Value} & \textbf{Dimension (2D)} \\
\hline
\done{Effective temperature} & $T^*$ & $[0, 0.4]$ & $1$ \\
Flow alignment & $\lambda$ & $[0, 1]$ & 1 \\
Rotational diffusivity & $\Gamma$ & $0.05$ & $T/M$ \\
Solvent viscosity & $\eta$ & $40/6$ & $M/T$ \\
Density & $\rho$ & $40$ & $M/L^2$ \\
%Friction & $\gamma$ & $1$, $0$ & $M/T$ (\jc{check}) \\
\done{Landau coefficient} & $A$ & $1$ & $M/T^2$ \\
Frank elastic constant & $K$ & $0.05$ & $ML^2/T^2$ \\
% \fix{Numerical integration time} & $\tau_{LB}$ & $1$ & $T$ \\
%Activity & $\zeta$ & $[0, 0.02]$ & $M/T^2$ \\
% Initial noise in alignment & $n_0$ & $0.05$ & 1 \\
% \fix{Director fluctuation} & $k_B T^Q$ & $[0, 0.05]$ & $ML^2/T^2$ \\
Square domain length & $L$ & $\{128, 256, 512\}$ & $L$ \\
\hline
\end{tabular}
}
\label{table:simulation_parameters}
\end{table}

\section{Results}

% MENTION: results are for L=256. with that one fig in appendix to show that results are ind of syystem size.
\done{We keep model parameters fixed while varying the effective temperature $T^*$ (occasionally referred to simply as \textit{temperature})}, which controls the strength of fluctuations in both $\mathbf{Q}$ and $\vec{u}$. 
\done{Gradually} increasing (decreasing) $T^*$ defines the forward (backward) protocol [Fig.~\ref{Fig1:unbinding_schematic}a].
\done{To achieve this gradual change in temperature,
the final frame for a simulation with temperature $T^*$ is used as the initial frame for $T^* + \Delta T^*$ and so on.}
After each change, the system is equilibrated before measurements are taken \done{(see App. \ref{app:hb} for additional details).}
\done{Finally, let us note that results in the main text are for $L=256$. In App. \ref{app:varying_systemsize}, we demonstrate that the main findings of this section are independent of system size.}

\subsection{Passive nematics without flow coupling}
To establish a baseline for the topological transition in the absence of hydrodynamic effects, we first analyze a purely relaxational nematic, in which the flow field is suppressed. With the flow field set to zero, \done{thermal} fluctuations act only on the orientational dynamics. Figure~\ref{Fig1:unbinding_schematic}(a--d) shows the expected BKT sequence. Above a creation threshold $T^*_{\text{dc}}\approx 0.13$, bound $\pm 1/2$ pairs appear, then separate as $T^*$ increases, yielding a gas of free disclinations (\done{see App. \ref{app:detecting_defects} for details on defect detection}).  Upon cooling, defects recombine and annihilate, restoring order. This reversible transition serves as our BKT benchmark.
%%%%%%%%%%%%%%%%%%%%%%%%%%%%%%%%%%%%%%%%%%%%%%%%%%%%%%%%%%%%%%%%%%%
\begin{figure*}[t!]
    \centering
\includegraphics[width=\linewidth]{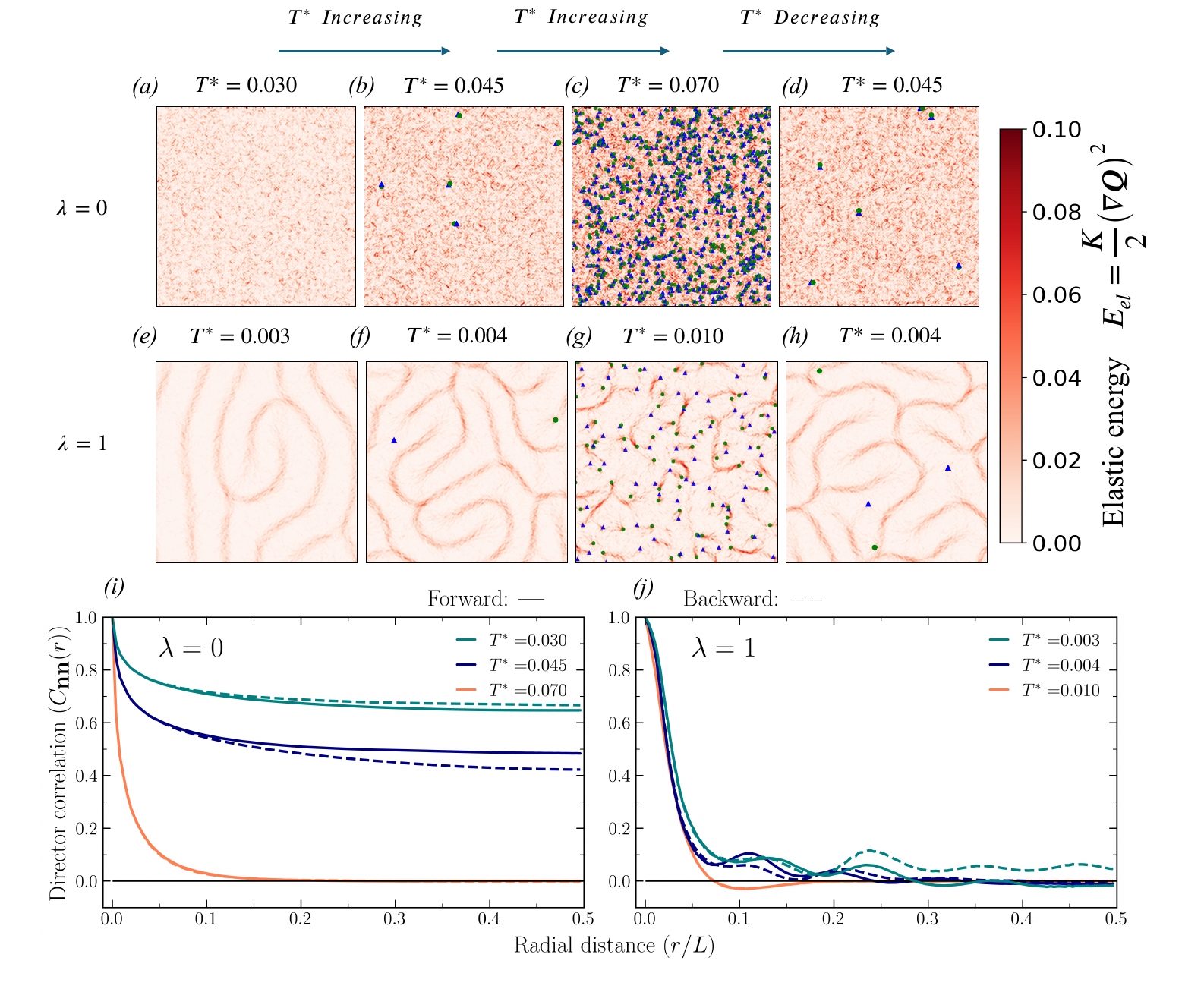}
    % \caption{Snapshots at different temperature with +1/2 defects as green circles, $-$1/2 defects as blue triangles, and the elastic energy as a heat map are shown in (a-d) for the non-aligning nematics with $\lambda = 0$  and for (e-h) strain-rate-aligning nematics with $\lambda = 1$. We note that bend walls are only present for $\lambda = 1$ and absent for $\lambda = 0$. 
    % (i,j) Averaged director-director correlations $C_{\textup{\textbf{n}\textbf{n}}}(r)$ for (i) $\lambda = 0$ and (j) $\lambda = 1$ for both protocols. For $\lambda=0$, in the forward protocol, quasi-long-range correlations in the director field are present when defects are absent ($T^*=0.030$) or bound ($T^* = 0.045$), whereas they decay \done{rapidly} when defects are unbound ($T^* = 0.070$), consistent with the expectation that defect unbinding destroys quasi-long-range order. In the backward case, this long-range order at low temperature is only partly restored, owing to the long relaxation time needed for full equilibration at low temperature. For $\lambda=1$, director correlations decay much more rapidly in comparison. Local maxima in the director correlations evidence the presence of bend wall and their characteristic spacing, which decreases with increasing temperature.
    % }
    \caption{\done{Snapshots of $+1/2$ (green circles) and $-1/2$ (blue triangles) defects and elastic-energy fields are shown for (a–d) non-aligning nematics ($\lambda=0$) and (e–h) strain-rate–aligning nematics ($\lambda=1$). Bend walls appear only for $\lambda=1$.
    Panels (i,j) show the averaged director–director correlations $C_{\mathbf{n}\mathbf{n}}(r)$ for
    the forward (solid lines) and backward (dashed lines) protocols. For $\lambda=0$, the forward protocol exhibits quasi-long-range correlations when defects are absent ($T^*=0.030$) or bound ($T^*=0.045$), with correlations decaying rapidly once defects are unbound ($T^*=0.070$). In the backward protocol, this trend is reversed, and defects rebind as the temperature is lowered. For $\lambda=1$, correlations decay much more rapidly, and local maxima reflect the presence of bend walls with a characteristic spacing that decreases as temperature increases.}}
    \label{Fig2:Dir-corr}
\end{figure*}
%%%%%%%%%%%%%%%%%%%%%%%%%%%%%%%%%%%%%%%%%%%%%%%%%%%%%%%%%%%%%%%%%%%
\subsection{Passive nematics with flow coupling}
Having studied the expected BKT binding–unbinding sequence in the flow–decoupled system, we now reintroduce hydrodynamics to determine how coupling to fluid motion modifies the transition. With hydrodynamics, behavior depends on flow alignment. We contrast (i) non-aligning nematics ($\lambda=0$) with (ii) strain-rate--aligning nematics ($\lambda\neq 0$). \done{We note that the BKT transition is violated for any regime where the director is coupled to strain flows, $\lambda \neq 0$, even for flow-tumbling regimes where there is no stable Leslie angle to the principal axis of flow deformation \cite{thijssen2020binding}.}

\subsubsection{Non-aligning nematics}
We first consider the non-aligning case, which isolates backflow from alignment effects and provides a direct comparison to the flow–decoupled benchmark. For $\lambda=0$, the system retains BKT-like behavior. The defect creation threshold is reduced ($T^*_{\text{dc}}\approx 0.042$), but the sequence remains: bound pairs at low $T^*$, unbinding at higher $T^*$, and recovery of order on cooling [Fig.~\ref{Fig1:unbinding_schematic}(e--h)]. Forward and backward protocols overlap within relaxation uncertainties, consistent with BKT expectations (see also Appendix. Fig.~\ref{Fig5:dry_rnn}).
\begin{figure*}[t!]
    \centering
    \includegraphics[width=1\linewidth]{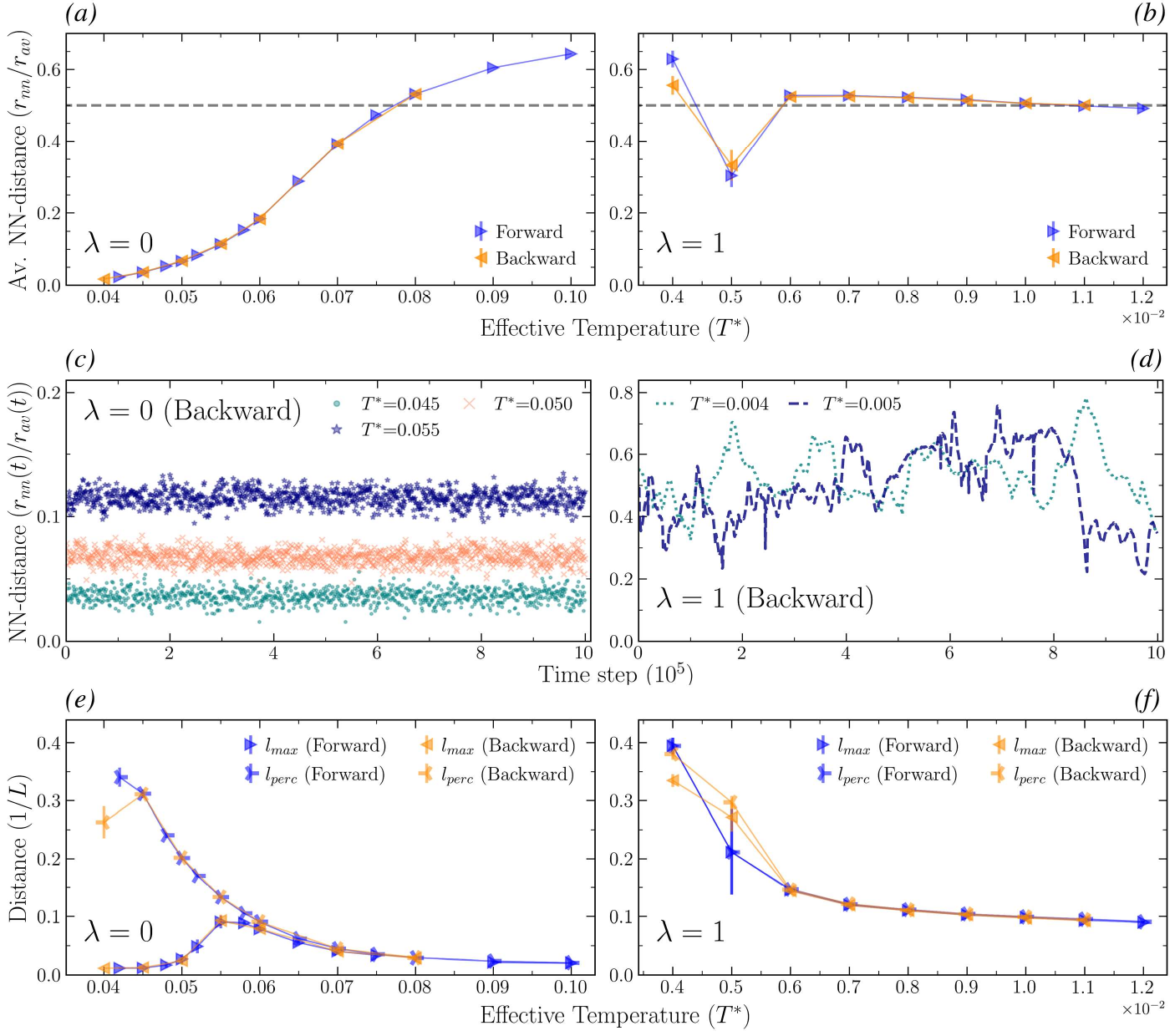}
% \caption{\done{Defect unbinding in non-aligning ($\lambda=0$) and strain-rate–aligning ($\lambda=1$) passive nematics.
% (a,b) Average nearest-neighbor distance $r_{nn}$, normalized by the average defect spacing $r_{av}$. The dashed line indicates the expectation for non-interacting defects, $r_{nn}^{\textup{free}} = 0.5 r_{av}$. For $\lambda=0$ (a), the forward protocol shows bound defects at low temperature ($r_{nn} \ll r_{nn}^{\textup{free}}$), with unbinding at higher temperatures as $r_{nn} \to r_{nn}^{\textup{free}}$. In the backward protocol, this trend is recovered for $T^* \ge 0.05$, whereas for $T^* < 0.05$ relaxation slows, causing $r_{nn}$ to increase. This slow relaxation is evident in (c), where $r_{nn}$ decreases only gradually at the lowest backward temperatures. For $\lambda=1$ (b), $r_{nn} \approx r_{nn}^{\textup{free}}$ at all temperatures, indicating unbound defects; correspondingly, (d) shows strong temporal fluctuations in $r_{nn}$ and discontinuous jumps from defect annihilation.
% (e,f) Cluster-based measures: the time-averaged neutralization length $l_{\max}$ and percolation length $l_{\text{perc}}$ for the forward protocol. For $\lambda=0$ (e), $l_{\max} < l_{\text{perc}}$ at low temperature (bound defects), while $l_{\max} = l_{\text{perc}}$ at higher temperature (unbound defects). For $\lambda=1$ (f), $l_{\max} = l_{\text{perc}}$ for all $T^*$, consistent with defects being unbound whenever present.}}
\caption{\done{Defect unbinding in non-aligning ($\lambda=0$) and strain-rate–aligning ($\lambda=1$) passive nematics.
(a,b) Average nearest-neighbor distance $r_{nn}$, normalized by the average defect spacing $r_{av}$. The dashed line indicates the expectation value for free and non-interacting defects, $r_{nn}^{\textup{free}} = 0.5 r_{av}$. For $\lambda=0$ (a), the forward protocol shows bound defects at low temperature ($r_{nn} \ll r_{nn}^{\textup{free}}$), with unbinding at higher temperatures as $r_{nn} \to r_{nn}^{\textup{free}}$. In the backward protocol, this trend is reversed, and defects rebind upon cooling. This rebinding is also evident from (c), which shows $r_{nn}(t)$ upon cooling for certain temperatures.
For $\lambda=1$ (b), $r_{nn} \approx r_{nn}^{\textup{free}}$ at all temperatures, indicating unbound defects; correspondingly, (d) shows strong temporal fluctuations in $r_{nn}(t)$
(with kinks indicating defect creation/annihilation).
(e,f) Time-averaged neutralization length $l_{\max}$ and percolation length $l_{\text{perc}}$ for both protocols. For $\lambda=0$ (e), $l_{\max} < l_{\text{perc}}$ at low temperature (bound defects), while $l_{\max} = l_{\text{perc}}$ at higher temperature (unbound defects). For $\lambda=1$ (f), $l_{\max} = l_{\text{perc}}$ for all $T^*$, consistent with unbound defects.}}
  \label{Fig3:defect_unbinding}
\end{figure*}

%%%%%%%%%%%%%%%%%%%%%%%%%%%%%%%%%%%%%%%%%%%%%%%%%%%%%%%%%%%%%%%%%%
\subsubsection{Strain-rate--aligning nematics: wall-mediated unbinding}
We now turn to the aligning regime. Introducing a nonzero flow–alignment parameter qualitatively reorganizes the director field and, as described below, changes the fate of defect pairs. For $\lambda\neq 0$, the dynamics depart from BKT. The defining feature is the spontaneous formation of bend--splay walls at
\done{non-zero temperatures $T^*$} [Fig.~\ref{Fig1:unbinding_schematic}(i)]. These extended orientational distortions partition the system into domains and dramatically lower the \done{threshold} for defect creation ($T^*_{\text{dc}}\approx 0.004$). Once defects appear, they do not form stable bound pairs: In both forward and backward protocols, defects are unbound whenever present
[Fig.~\ref{Fig1:unbinding_schematic}(j--l)]. 
%\sg{ \sout{Once defect pairs appear \kt{Comment Twice said "once defect appears"}, they unbind immediately: the $+1/2$ defects are advected throughout the system, while the $-1/2$ defects remain almost immobile. As the temperature increases, the system becomes progressively populated by free defects. Upon lowering the temperature, the same path is retraced, so defects remain unbound at all $T^*$ in both the forward and backward protocols.}  \textbf{[Is any of this needed?]}}

\paragraph*{Mechanism: why walls imply persistent unbinding.}
The co-rotation term couples shear to orientation for $\lambda\neq 0$. 
\done{Thermal fluctuations generate strain rates that locally couple the director to strain rates, often called elastic backflow. This in turn causes persistent flow fields around long-range deformations and topological defects \cite{krommydas2023hydrodynamic, Lasse-PRE}. Gradients of these persistent flows cause spatial variations in this alignment, which focus bend/splay distortions into narrow walls (visible as elastic-energy ridges in Fig.~\ref{Fig1:unbinding_schematic}j--l) as was analytically shown to occur \cite{zhu2020investigation}. The distinct behavior for $\lambda\neq 0$ can be traced to the spontaneous formation of bend–splay walls, which is in contrast to the case of $\lambda= 0$, where no persistent flow structures appear around orientation deformations \cite{krommydas2023hydrodynamic, Lasse-PRE}. In addition,  this elastic backflow causes an asymmetry between $+1/2$ and $-1/2$ defects, which also promotes unbinding \cite{krommydas2023hydrodynamic, Lasse-PRE}. 
\\
We summarize the resulting mechanism for  $\lambda\neq 0$ in three components: (i) Nucleation facilitation: Curvature and splay concentrate along walls due to elastic backflow, resulting in persistent flows which reduce the elastic cost of introducing defect cores; wall ends and junctions act as preferred nucleation sites, lowering $T^*_{\textup{dc}}$. (ii) Redirection: Across a wall, the director rotates by a finite angle, segmenting space into domains with misaligned orientations. The effective “Coulombic” (logarithmic) interaction between $\pm1/2$ charges is then redirected by intervening walls, so that approaching opposite charges are diverted along walls rather than drawn together across them. (iii) Guided transport without pairing: Flow gradients are strongest near walls; because the alignment feedback maintains these gradients, defects are advected along walls and toward junctions where many defects circulate and exchange partners. This promotes mixing of charges but not long-lived pair stabilization. These last two effects are in \done{agreement} to recent work in active nematic systems, where wall-like structures induce an apparent arrest of defects \cite{lavi2024dynamical}.}

These structural signatures are evident directly in the elastic energy heat maps of the director field, as well as a radial director-director correlation function defined as $C_{\textup{\textbf{n}\textbf{n}}}(r) = \frac{\langle |\textbf{n} (0) \textbf{n} (\textbf{r})| - 2/\pi \rangle }{ \langle | \textbf{n} (0) \textbf{n} (0)| - 2/\pi \rangle}$. $\langle \cdot \rangle$ denotes averaging over space, time and orientation, and $2/\pi$ is subtracted to ensure that $C_{\textup{\textbf{n}\textbf{n}}}(r)$ vanishes as $r\rightarrow \infty$ in the absence of correlations.

Figure~\ref{Fig2:Dir-corr} provides a detailed view: in the aligning case, elastic energy maps reveal the growth of bend--splay walls [Fig. ~\ref{Fig2:Dir-corr} e--h], while correlation functions show damped oscillations with multiple local maxima [Fig. ~\ref{Fig2:Dir-corr}j], indicating finite-size domains. By contrast, in the non-aligning case, director correlations display quasi--long-range order at low \done{temperature} [Fig.~\ref{Fig2:Dir-corr}i], which decays sharply only once defects unbind. 

\done{In the backward protocol (dotted lines), defects rebind as the temperature is lowered, and quasi--long-range order is restored. Such reversibility is expected for the BKT-transition, and is also observed in our BKT benchmark system: the flow-decoupled nematic [see App.Fig.~\ref{Fig5:dry_rnn}b]. 
In contrast, for aligning nematics, 
the system reverts to its equilibrium state only stochastically upon cooling, in the sense that 
defect interactions are screened by the presence of walls.
Consequently, defect annihilation - and thus relaxation toward equilibrium - is contingent on defects randomly recombining.}

\subsection{Quantifying defect unbinding}
The differences between non-aligning and aligning systems can be quantified using defect statistics. We employ two complementary approaches: (i) the nearest-neighbor distance \(r_{nn}\), \done{which we compare to the expectation value of the nearest-neighbor probability distribution for uncorrelated and uniformly distributed points: $r_{nn}^{\textup{free}} = 0.5 ~r_{av}$ ~\cite{nieves_preprint_2024}, where the average distance between defects $r_{av}=1/\sqrt{\rho_D}$ is set by the defect number density $\rho_D$.} (ii) a defect clustering scheme~\cite{Nuno-PhysRevE.65.066117}, in which two defects are assigned to the same cluster if their separation is less than a specified cutoff distance $l_{\textup{cutoff}}$. For each $l_{\textup{cutoff}} \in \{1,2, ..., L/2\}$, the total topological charge of each cluster is recorded, and the \textit{neutralization length} $l_{\textup{max}}(t)$ at time $t$ is then defined as the cutoff distance, at which all clusters are neutral. Similarly, the \textit{percolation length} $l_{\textup{perc}}(t)$ is defined as the cutoff distance, at which all defects belong to the same cluster. When defects are tightly bound, all clusters become neutral when the cutoff length equals the separation of the defect pair that is most loosely bound. For systems of bound defects, therefore, $l_{\textup{max}}(t)$ measures the maximum separation among all defect pairs, and $l_{\textup{max}}(t) < l_{\textup{perc}}(t)$. Conversely, when defects are unbound, all clusters will not generally become neutral until the cutoff length is large enough that only one cluster remains, i.e. when $l_{\textup{max}}(t) = l_{\textup{perc}}(t)$ (for more details, see~\cite{Nuno-PhysRevE.65.066117}).

Figure~\ref{Fig3:defect_unbinding}(a) shows that for non-aligning nematics, \(r_{nn}\) is significantly smaller than the free-defect expectation $r_{nn}^{\textup{free}}$ at low temperature $T^*$, indicating tightly bound pairs. As \done{temperature is increased}, \(r_{nn}\) approaches $r_{nn}^{\textup{free}}$, signaling unbinding.
\done{
In the backward protocol, this trend is reversed, with free defects becoming bound as the temperature is lowered.
This rebinding is also evident from (c), which shows that after cooling, the nearest-neighbor distance $r_{nn}(t)$ is small and roughly constant over time.}
These findings match those of the flow-decoupled nematic \done{[Fig.~\ref{Fig5:dry_rnn}(c,d) in App. \ref{app:dry_passive}].}

For strain-rate--aligning nematics, the outcome is different. Here, \(r_{nn}\) remains close to the free-defect expectation across all temperatures and both protocols [Fig.~\ref{Fig3:defect_unbinding}b], and time traces reveal significant variability in the nearest-neighbor distance, even at the lowest temperatures [Fig.~\ref{Fig3:defect_unbinding}d], indicating a rough energy landscape that is only slowly explored through thermal fluctuations. The defect clustering analysis confirms these trends. Letting $l_{\max}$ denote the time average of $l_{\max}(t)$ (and similarly for $l_{\text{perc}}$), we see that in the non-aligning case, $l_{\max} \ll l_{\textup{perc}}$ at low temperature, reflecting tightly bound defect pairs. As temperature increases, the two curves merge, consistent with unbinding [Fig.~\ref{Fig3:defect_unbinding}e]. 
\done{As revealed by the backward protocol, lowering the temperature reverses this process.} Once again, these results echo those of the flow-decoupled nematic \done{[Fig.~\ref{Fig5:dry_rnn}a in App. \ref{app:dry_passive}]}. In the aligning case, on the other hand, \(l_{\max}\) and \(l_{\text{perc}}\) coincide for all temperatures [Fig.~\ref{Fig3:defect_unbinding}f], showing that defects never form stable bound pairs.

Finally, let us emphasize that the unbinding behavior of defects and its temperature dependence is independent of system size \done{[Fig.~\ref{Fig6:rnn_vs_L} in App. \ref{app:varying_systemsize}]}.

\subsection{\done{Active system}}
\label{Results: Active nematics}
\begin{figure*}[t!]
    \centering
\includegraphics[width=0.85\linewidth]{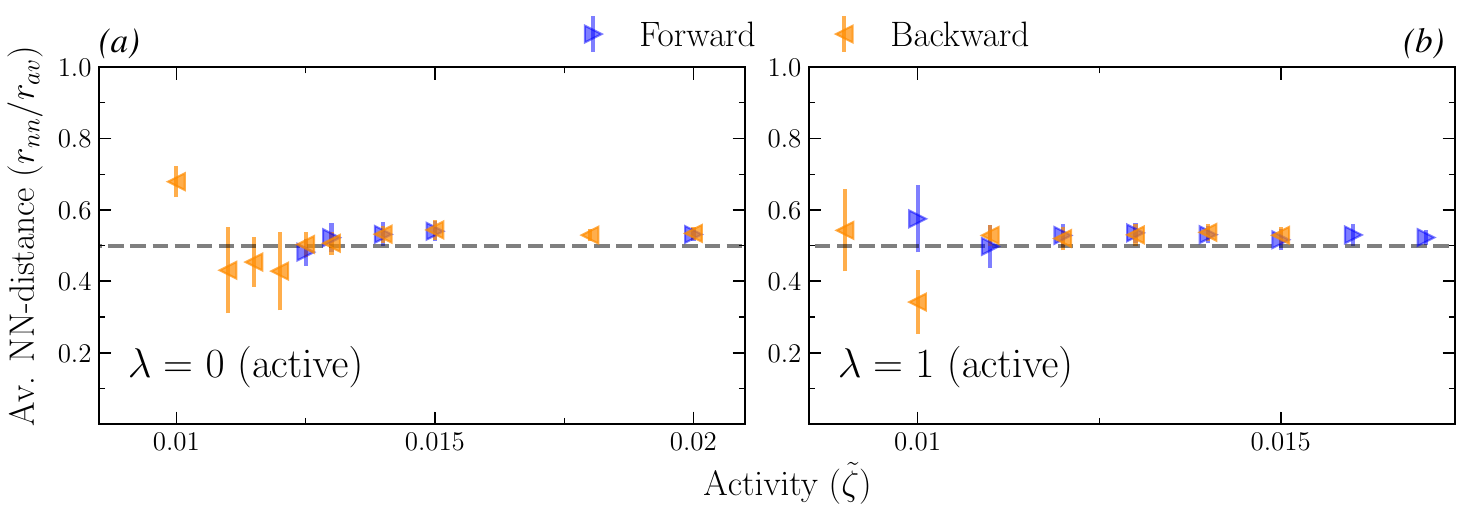}
    \caption{\done{Average nearest-neighbor distance, $r_{nn}$, for active, 
    wet nematic systems with alignment parameters $\lambda=0,1$ against 
    activity, $\tilde{\zeta} = \zeta / A$, where $A$ is the Landau–de Gennes 
    coefficient. Results are expressed in units of the average defect–defect 
    distance $r_{av}$, and the dotted horizontal line marks the expected 
    nearest-neighbor distance for independently distributed points, 
    $r_{nn}^{\textup{free}} = 0.5 r_{av}$. 
    (a) $\lambda=0$: In contrast to the passive case 
    [Fig.~\ref{Fig3:defect_unbinding}(a)], the nearest-neighbor distance remains close 
    to $r_{nn}^{\textup{free}}$ across all activities, showing that defects 
    are always unbound. 
    (b) $\lambda=1$: As in the $\lambda=0$ case, defects are unbound 
    whenever present. 
    }}
    \label{Fig4:rnn_active}
\end{figure*}

\done{
%In active nematic system, self-generated stresses drive continuous defect creation and motility.
We next consider active nematics, where activity contributes an additional stress, $\Pi^{active}_{ij}= -\zeta Q_{ij}$, proportional to the local nematic order \cite{doostmohammadi2018active, thampi_active_2016}. 
Unlike thermal noise, activity continuously injects energy and sustains a steady population of defects.
%To quantify defect binding, we compute the average nearest-neighbor distance $r_{nn}$ for both $\lambda=0$ and $\lambda=1$.  
Figures~\ref{Fig4:rnn_active}(a,b) show $r_{nn}$ normalized by the average defect spacing $r_{av}$ across a range of activity strengths $\tilde{\zeta} = \zeta / A$.
\revtwo{Here, we focus only on the case of positive $\zeta$, corresponding to extensile active stresses, since activity-induced nematic order in contractile systems could involve additional mechanisms~\cite{venkatesh2026emergent}.}
For both $\lambda=0$ and $\lambda=1$, $r_{nn}$ remains close to the 
free-defect value $r_{nn}^{\textup{free}} = 0.5 \, r_{av}$ over the entire 
activity range, and this holds in both forward and backward protocols. 
This demonstrates that in active systems defects are unbound whenever 
present, independent of the alignment parameter.  
Even at the lowest activities accessible in our simulations, defect pairs 
do not remain bound but instead separate quickly, and $r_{nn}(t)$ fluctuates 
around the free-defect expectation, consistent with persistent motility of $+1/2$ defects.} 

\done{This behavior is striking when contrasted with the passive friction-dominated 
case: for $\lambda=0$ in equilibrium, we observed a clear binding–unbinding 
transition consistent with the BKT mechanism. This is because there were no persistent flow gradients, which could result in wall formation. However, in an active nematic system, we find persistent flow structures even for $\lambda=0$.
Activity therefore removes the 
possibility of defect binding altogether through the previous argued mechanism, driving the system into a regime 
of sustained defect unbinding regardless of $\lambda$. 
These results emphasize that fluid flow—whether generated intrinsically by 
activity or coupled viscously—suppresses the equilibrium binding mechanism 
and sustains turbulent defect dynamics.}
%%%%%%%%%%%%%%%%%%%%%%%%%%%%%%%%%%%%%%%%%%%%%%%%%%%%%%%%%%%%%%%%%%
%%%%%%%%%%%%%%%%%%%%%%%%%%%%%%%%%%%%%%%%%%%%%%%%%%%%%%%%%%%%%%%%%%

\section{Discussion}
Coupling nematic order to flow fundamentally reshapes defect-mediated transitions. The decisive control parameter is the flow--alignment coefficient $\lambda$. For non-aligning nematics ($\lambda=0$), the classical BKT binding--unbinding mechanism is retained. For nonzero alignment ($\lambda\neq 0$), bend--splay walls emerge; they lower defect nucleation thresholds and, via screening and guided transport along wall networks, prevent sustained recombination. Consequently, once created, defects remain unbound across \done{temperatures} and thermal protocols \done{due to the velocity asymmetry of $+1/2$ and $-1/2$ defects found in ($\lambda\neq 0$) systems} \cite{krommydas2023hydrodynamic, Lasse-PRE}.

This extends the classical rheology of flow alignment \cite{de1993physics, beris_thermodynamics_1994} into the realm of topological phase behavior: flow alignment does not merely renormalize defect mobilities; it determines whether bound states are viable. Since most nematics possess finite alignment, our results suggest that the canonical BKT transition is realized only in the absence of flow alignment ($\lambda = 0$). 

\done{In active nematics, self-sustained flow provides a continuous source of defects, and once created they likewise remain unbound irrespective of alignment. Together, these results identify flow coupling as a fundamental control parameter for defect-mediated transitions, demonstrating how hydrodynamic and active flows reshape topological phase behavior in nematic fluids.}

These predictions invite experimental tests in colloidal and molecular nematics and in thin films under controlled shear/strain fluctuations. Direct measurements of defect pair statistics together with imaging of wall networks would critically assess the wall-mediated mechanism. Theoretically, extensions to three dimensions—where line defects, biaxiality, and secondary instabilities add richness—may reveal new classes of flow-driven topological transitions. 

%%%%%%%%%%%%%%%%%%%%%%%%%%%%%%%%%%%%%%%%%%%%%%%%%%%%%%%%%%%%%%%%%%
\section*{Acknowledgements}
A. D. acknowledges funding from the Novo Nordisk Foundation (grant No. NNF18SA0035142 and NERD grant No. NNF21OC0068687), Villum Fonden (Grant no. 29476), and the European Union (ERC, PhysCoMeT, 101041418). K.T acknowledges funding from the Novo Nordisk Foundation. (grant No.
NNF23OC0085012). Views and opinions expressed are however those of the authors only and do not necessarily reflect those of the European Union or the European Research Council. Neither the European Union nor the granting authority can be held responsible for them. The Tycho supercomputer hosted at the SCIENCE HPC center at the University of Copenhagen was used for supporting this work. 

%%%%%%%%%%%%%%%%%%%%%%%%%%%%%%%%%%%%%%%%%%%%%%%%%%%%%%%%%%%%%%%%%%
\section*{Data Availability}
\done{The data that support the findings of this study
will be openly available following an embargo at
the following URL/DOI: TBA.}

\appendix

\section{\done{Simulation details}}
\label{app:hb}

\noindent
\done{The governing Eqs.  \eqref{eq-Q},  \eqref{eq-u} are solved using the hybrid Lattice Boltzmann method developed by Marenduzzo et al. \cite{Marenduzzo_hlb_2007}. The evolution of the flow velocity $\vec{u}$ is determined by solving the generalized Navier-Stokes Eqs. \eqref{eq-u} using the Lattice Boltzmann method (LBM), while the time evolution of nematic order parameter $\mathbf{Q}$ is found by solving the Beris-Edwards Eq. \eqref{eq-Q} using a
five-point stencil approximation. For the LBM, we have used a $D2Q9$ grid for discretizing the velocity, and the so-called BKG collision operator to model the collision term in the Boltzmann equation \cite{bhatnagar_bkg1954}, with a relaxation time equal to the simulation time step.}

\done{Simulations are performed 
with simulation time step $\Delta t = 1$ and lattice spacing $\Delta x=1$ on square lattices with side length $L\in\{128, 256, 512\}$ and periodic boundary conditions. Each run is equilibrated for $1.5\cdot10^{5}$ time steps (for low temperatures, $5\cdot10^6$ steps are needed to ensure proper equilibration), followed by $10^{6}$ time steps for data collection. Frames are stored every 1000 time steps, yielding a total of 1000 samples. \revtwo{The thermodynamic quantities (average defect density, average nearest-neighbor distance, etc.) are calculated by averaging over 1000 samples.}
In the forward (backward) protocol, 
temperature is raised (decreased) gradually as follows:
the final frame for a simulation with temperature $T^*$ is used as the initial frame for $T^* + \Delta T^*$ and so on.}

\section{\done{Detecting defects}}
\label{app:detecting_defects}

\done{Defects are found by calculating the winding number of the director, specifically, by employing the algorithm by Zapotocky et al. as follows \cite{zapotocky_kinetics_1995}: For every square of $2\times2$ neighboring lattice points in the system, the cumulative change in the orientation of the director is calculated as it moves around the square in the counterclockwise direction. A directional change of $\pi$ ($-\pi$) indicates the presence of a $+1/2$ ($-1/2$) defect at the center of the square.}

\vspace{2em}

\section{\done{Results for} passive nematics without flow coupling}
\label{app:dry_passive}
\begin{figure*}[t!]
    \centering
\includegraphics[width=1\linewidth]{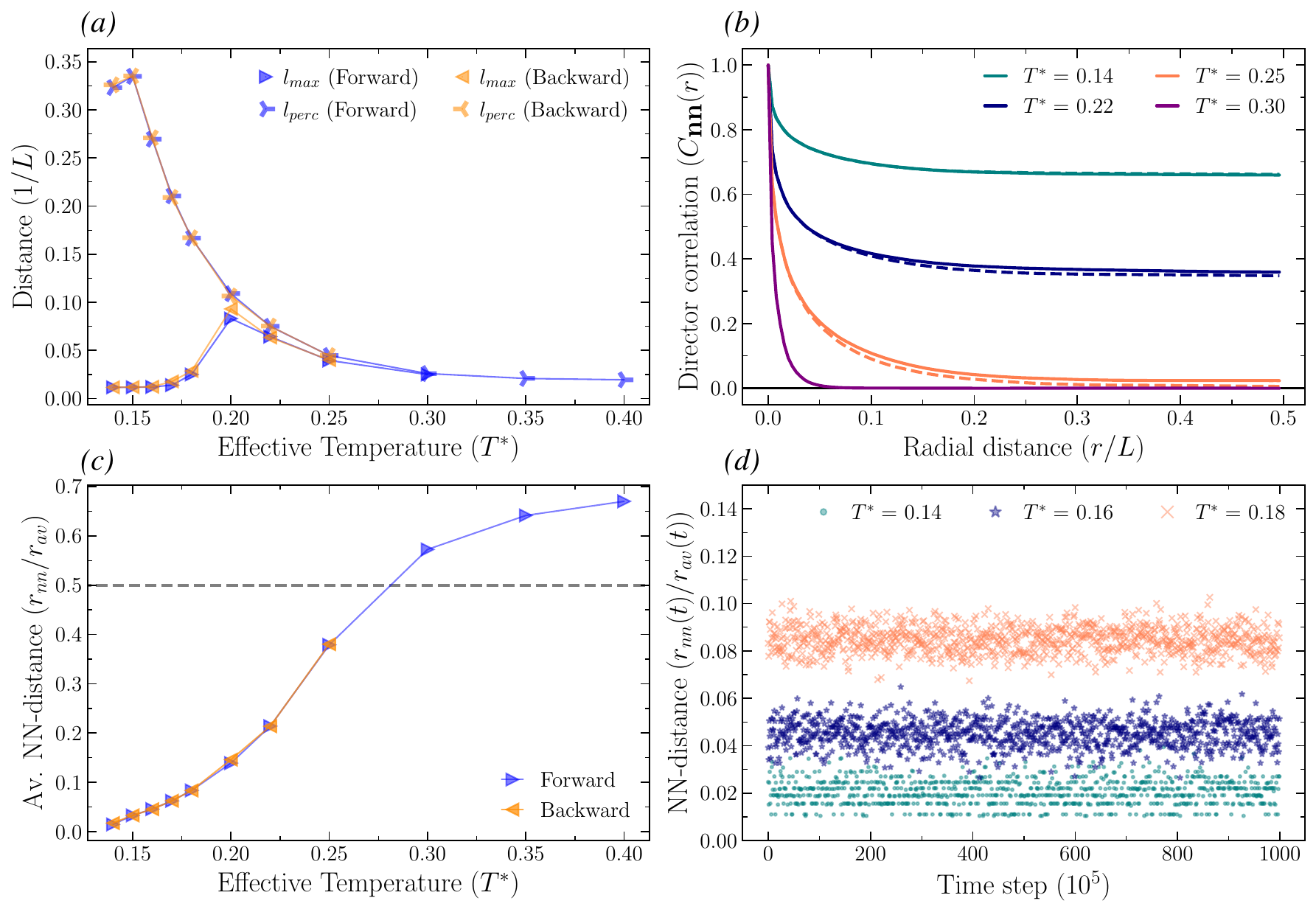}
\caption{\done{Defect unbinding and orientational order in passive nematics without flow.
(a) Time-averaged neutralization length $l_{\max}$ and percolation length $l_{\text{perc}}$ for forward and backward protocols. 
In both protocols, $l_{\max} < l_{\text{perc}}$ at low temperature (bound defects), while $l_{\max} = l_{\text{perc}}$ at higher temperature (unbound defects).
(b) Director correlation function $C_{\mathbf{n}\mathbf{n}}(r)$ for representative temperatures, with solid (dashed) curves for the forward (backward) protocol. At $T^*=0.14$, tightly bound defects yield quasi–long-range order; at $T^*=0.22$, defects are more loosely bound, and the order weakens. For $T^* \ge 0.25$, 
correlations decay rapidly, consistent with defect unbinding. 
In the backward protocol, this trend is reversed.
(c) Average nearest-neighbor distance $r_{nn}$ normalized by the average defect spacing $r_{av}$. The dashed line indicates the expectation value for free defects, $r_{nn}^{\textup{free}} = 0.5 r_{av}$.
The forward protocol shows bound defects at low temperature ($r_{nn} \ll r_{nn}^{\textup{free}}$), with unbinding at higher temperatures as $r_{nn} \to r_{nn}^{\textup{free}}$. In the backward protocol, this trend is reversed, and defects rebind as the temperature is lowered. Such rebinding is evident from (d), which shows that upon cooling, $r_{nn}(t)$ is small and roughly constant.}}
\label{Fig5:dry_rnn}
\end{figure*}

In passive nematics without flow coupling, defect statistics reveal a binding–unbinding transition consistent with the BKT scenario. The cluster-based measures $l_{\max}$ and $l_{\text{perc}}$ [Fig.~\ref{Fig5:dry_rnn}(a)] show that defects remain bound for $T^* \lesssim 0.20$, as indicated by $l_{\max} < l_{\text{perc}}$. At higher temperatures, $l_{\max}$ converges to $l_{\text{perc}}$ within uncertainty for $T^* \gtrsim 0.25$, indicating the onset of defect unbinding. These results are consistent with the behavior of the director correlation function $C_{nn}(r)$ [Fig.~\ref{Fig5:dry_rnn}(b)], which exhibits quasi–long-range order for $T^* = 0.15$ when defects are tightly bound, weakened correlations at $T^* = 0.22$ when defects are more loosely bound, and a rapid decay for $T^* \geq 0.25$ where unbound defects proliferate. Nematic order is partially restored in the backward protocol which is due to the slow relaxation of the defects. The nearest-neighbor analysis [Fig.~\ref{Fig5:dry_rnn}(c)] further supports this interpretation: for $T^* \lesssim 0.22$, $r_{nn}$ lies well below the free-defect value $r_{nn}^{\textup{free}}$, indicating bound pairs, while at larger temperatures $r_{nn}$ approaches $r_{nn}^{\textup{free}}$, consistent with unbound defects. In the backward protocol, however, for $T^* < 0.17$, $r_{nn}$ is elevated relative to equilibrium expectations due to the slow relaxation of defect pairs at low temperatures. This is illustrated in Fig.~\ref{Fig5:dry_rnn}(d), where $r_{nn}(t)$ decreases only slowly with time, and discontinuous drops mark defect–pair annihilation events. Together, these complementary measures establish that in the absence of flow coupling, passive nematics undergo a BKT-like binding–unbinding transition, with slow relaxation effects influencing the backward path at low temperatures.

\section{\done{Results across system sizes}}
\label{app:varying_systemsize}
\begin{figure*}[t!]
    \centering
\includegraphics[width=1\linewidth]{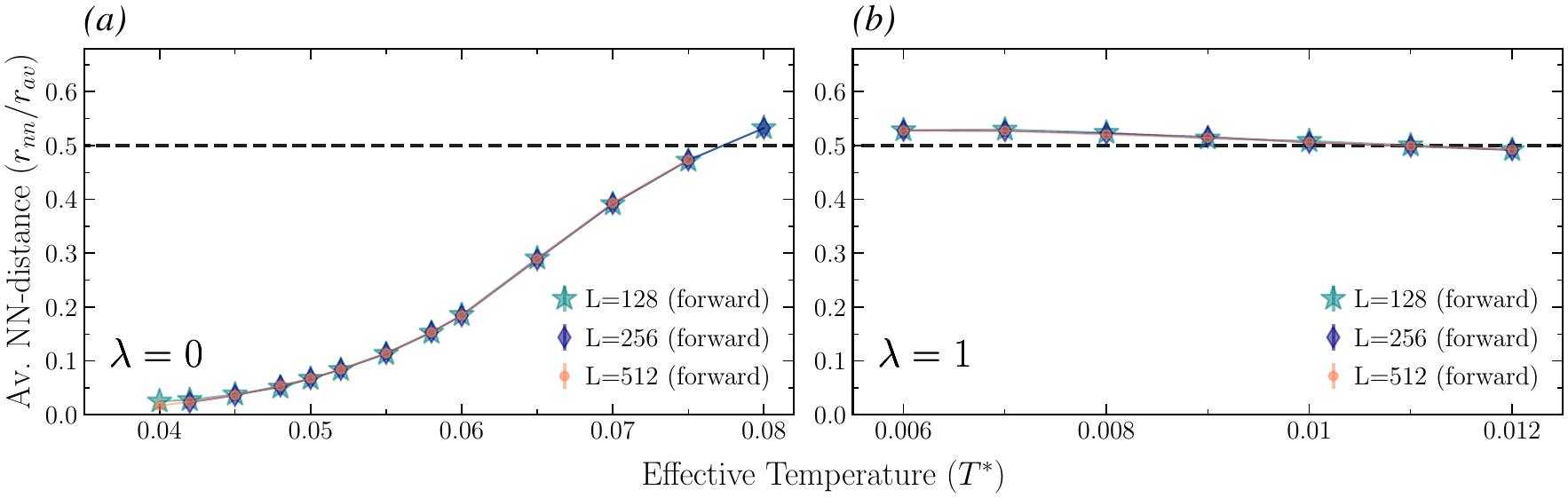}
\caption{\done{Varying system size: Average nearest-neighbor distance, $r_{nn}$, for non-aligning ($\lambda=0$) and strain-rate–aligning ($\lambda=1$) passive nematics
    for simulations with side length $L\in\{128,256,512\}$. 
    %To avoid clutter, 
    For visibility,
    only the forward protocols are included. Results are expressed in units of the average defect–defect distance $r_{av}$, and the dotted horizontal line marks the expected 
    nearest-neighbor distance for independently distributed points, $r_{nn}^{\textup{free}} = 0.5 r_{av}$. For $\lambda=0$ (a), defects are bound at low temperature but unbind at higher temperatures. For $\lambda=1$ (b), 
    $r_{nn} \approx r_{nn}^{\textup{free}}$ at all temperatures, signifying that defects are unbound whenever present. In both cases, 
    unbinding behavior and its temperature dependence
    is independent of system size.}}
  
    \label{Fig6:rnn_vs_L}
\end{figure*}

\done{
This appendix is included to show that the unbinding behavior of defects and its temperature dependence is independent of the simulation system size. This is evident from Fig. \ref{Fig6:rnn_vs_L}, which shows the average nearest neighbor distance for for non-aligning ($\lambda=0$) and strain-rate–aligning ($\lambda=1$) passive nematics simulations with side lengths $L\in\{128,256,512\}$. 
For visibility, only the forward protocols are included. Indeed, we see that---across system sizes---when $\lambda=0$, defects are bound at low temperature and unbind at higher temperature, whereas when $\lambda = 1$, defects are unbound whenever present.}

\clearpage
\bibliography{Ref}% Produces the bibliography via BibTeX.

\end{document}